\documentclass[fleqn,twoside]{article}
\usepackage{espcrc2}
\usepackage{amssymb}
\usepackage[dvips]{graphicx}

\newcommand{\AmS}{{\protect\the\textfont2
  A\kern-.1667em\lower.5ex\hbox{M}\kern-.125emS}}
\newcommand{\ped}[1]{\ensuremath{_{\rm #1}}}

\title{Point-contact spectroscopy in MgB$\ped{2}$ single
        crystals in magnetic field}

\author{D. Daghero\address{INFM - Dipartimento di Fisica, Politecnico di
Torino,
        Corso Duca degli Abruzzi 24, 10129 Torino, Italy}%
        \thanks{Corresponding author. e-mail: ddaghero@polito.it},
        R.S. Gonnelli\addressmark,
        G.A. Ummarino\addressmark,
        V.A. Stepanov\address{P.N. Lebedev Physical Institute, Russian Academy
        of Sciences, Leninski Pr. 53, 119991 Moscow, Russia},
        J. Jun\address{Solid State Physics Laboratory, ETH, CH-8093 Z\"{u}rich,
        Switzerland},
        S.M. Kazakov\addressmark,
        and
        J. Karpinski\addressmark}

\begin{document}

\begin{abstract}
We present the results of a spectroscopic study of
state-of-the-art MgB$\ped{2}$ single crystals, carried out by
using a modified point-contact technique. The use of single
crystals allowed us to obtain point contacts with current
injection either parallel or perpendicular to the $ab$ planes. The
effect of magnetic fields up to 9 T on the conductance spectra of
these contacts is here thoroughly studied, for both $\mathbf{B}$
parallel and perpendicular to the $ab$ planes. The complete
thermal evolution of the upper critical field of the $\pi$ band is
determined for the first time, and quantitative information about
the upper critical field of the $\sigma$ band and its anisotropy
is obtained.  Finally, by exploiting the different effect of a
magnetic field applied parallel to the $ab$ planes on the two band
systems, the partial contributions of the $\sigma$ and $\pi$ bands
to the total conductance are obtained separately. Fitting each of
them with the standard BTK model yields a great reduction of the
uncertainty on $\Delta\ped{\sigma}$ and $\Delta\ped{\pi}$, whose
complete temperature dependence is obtained with the greatest
accuracy.
\end{abstract}

\maketitle

\section{Introduction}
About two years after the discovery of superconductivity in
MgB$_2$, the validity of the two-gap model
\cite{Suhl,Liu,Brinkman,Mazin} in describing the superconducting
and normal-state properties of this material has been confirmed
by a large number of experimental evidences. Within this model,
the complex electronic bandstructure of MgB$_2$
\cite{Kortus,Shulga,An,Kong} is approximated by one quasi-2D
$\sigma$ band and one 3D $\pi$ band, featuring two gaps of very
different amplitude, $\Delta\ped{\sigma}$ and $\Delta\ped{\pi}$.
The different spatial character of the two bands arises from the
layered structure of the material, that is also expected to give
rise to macroscopic anisotropy in some relevant physical
quantities, e.g. penetration depth, coherence length, and upper
critical fields. The value of the anisotropy ratio
$\gamma=H\ped{c2}^{ab}/H\ped{c2}^{c}$ in MgB$_2$ has long been a
matter of debate, because of the large spread of values measured
in films and polycrystalline samples \cite{Buzea}. To this
regard, the recent set-up of efficient crystal-growth techniques
has thus been an essential improvement. As a matter of fact,
torque magnetometry \cite{Angst} and thermal conductivity
measurements \cite{Sologubenko} in high-quality single crystals
have given substantially consistent values of the upper critical
fields as a function of the temperature.

In this paper, we present the results of directional point-contact
spectroscopy (PCS) in state-of-the-art single crystals, in the
presence of a magnetic field applied either parallel or
perpendicular to the $ab$ planes. The use of PCS, and the
possibility of controlling the directions of both the injected
current and the field, allowed us to study the effect of the
magnetic field on each band system, separately. It turns out that
the magnetic field strongly affects the superconductivity in the
$\pi$-band, irrespective of its orientation. An analysis of the
conductance curves measured at different temperatures and various
field intensities gives the temperature evolution of the
$\pi$-band upper critical field, $B\ped{c2}^{\pi}=\mu_0
H\ped{c2}^{\pi}$, determined here for the first time. In contrast,
the effect of the field on the $\sigma$ band is highly
anisotropic. Our measurements indicate that the upper critical
field of the $\sigma$ band for $\mathbf{B}$ parallel to the $c$
axis, $B\ped{c2
\parallel c}^{\sigma}$, is higher than that measured by other
groups on similar samples \cite{Angst,Sologubenko}, possibly
because of surface effects. Finally, the temperature dependence of
the two gaps is obtained with unprecedented accuracy, by
separating the partial $\sigma$- and $\pi$-band contributions to
the total conductance by means of a suitable magnetic field.

\section{Experimental set up}
All the details about the sample preparation are given in the
paper by Karpinski \emph{et al.} in this same issue. The MgB$_2$
single crystals we used for our measurements were about $0.6\times
0.6 \times 0.04 $~mm$^3$ in size, even though the growth technique
allows obtaining even larger samples. The crystals were etched
with 1\% HCl in dry ethanol, to remove possible deteriorated
surface layers. The critical temperature of the crystals, measured
by AC susceptibility, is $T\ped{c}= 38.2$ K with $\Delta
T\ped{c}\sim 0.2$ K.

Possibly because of the extreme hardness of the crystals, point
contacts obtained by pressing a metallic tip against the crystal
surface were found to fail the essential requirements of
reproducibility and mechanical stability upon thermal cycling.
Therefore, we made the contacts by using as a counterelectrode
either a small drop of silver conductive paint ($\varnothing
\lesssim 50 \mu$m) or a small spot of indium. The control of the
junction characteristics (which, in the conventional technique, is
obtained by moving the tip) was a little more difficult in this
case. However, by applying short voltage pulses we were able to
change the characteristics of the contact, until satisfactory
stability and reproducibility were attained.

Of course, the apparent size of our contacts is much greater than
that required to have ballistic transport across the junction
\cite{ballistic}. Actually, the effective electrical contact
occurs via parallel micro-bridges in the spot area, that can be
thought of as Sharvin contacts. This assumption is supported
\emph{a posteriori} by the absence of heating effects in the
conductance curves of all our junctions. Moreover, by using in
the Sharvin formula the experimental resistance of our contacts
(that always fell in the range $10\div 50\, \Omega$) the
estimated mean free path $\ell= 600$\AA, and the residual
resistivity $\rho\ped{0}\approx 2 \mu \Omega\,$cm
\cite{Sologubenko} one obtains that, at least in the
higher-resistance junctions, the transport is ballistic even if a
single contact is established.

The contacts were positioned on the crystal surfaces so as to
inject the current along the $c$ axis or along the $ab$ planes.
In the following, we will refer to them as ``$c$-axis contacts''
and ``$ab$-plane contacts''. Notice that, when the potential
barrier at the interface is small as in our case, the current is
injected in a cone whose angle is not negligible, and becomes
equal to $\pi/2$ in the ideal case of no barrier. Anyway, the
probability for electrons to be injected along an angle $\phi$ in
the cone is proportional to $\cos{\phi}$ \cite{Tanaka} so that the
(main) direction of current injection can still be defined.

\section{Experimental results and discussion}
\subsection{Magnetic field parallel to the $c$ axis}
\vspace{3pt}

\begin{figure}[t]
\begin{center}
\includegraphics[keepaspectratio, width=0.9\columnwidth]{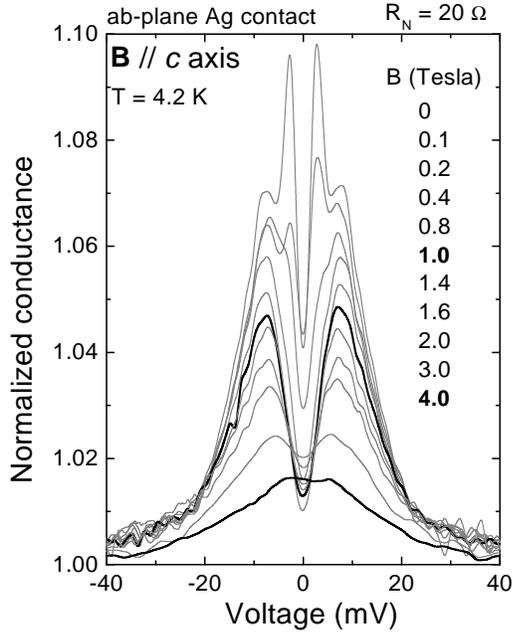}
\end{center}
\vspace{-2cm} \caption{\small{Effect of a magnetic field of
increasing intensity, applied parallel to the $c$ axis, on the
conductance curves of a Ag/MgB$_2$ point contact with current
injection mainly along the $ab$ plane. All the curves were
measured at $T=4.2$~K. Thick black lines represent the
experimental curves at
$B=1$~T and $B=4$~T.}} \label{fig:1}%\vspace{-1cm}
\end{figure}
\begin{figure}[t]
\vspace{-3mm}
\begin{center}
\includegraphics[keepaspectratio, width=0.9\columnwidth]{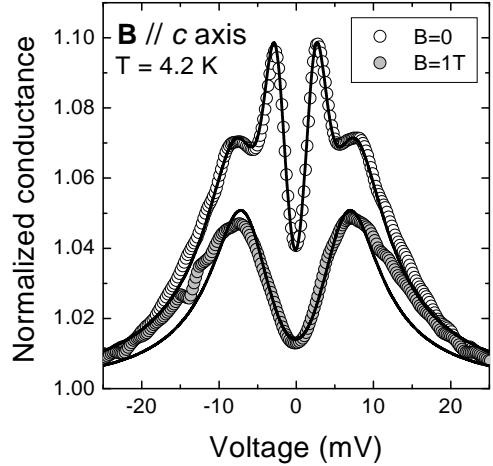}
\end{center}
\vspace{-3.5cm} \caption{\small{The experimental normalized
conductance curves at $B=0$ (open symbols) and $B=1$~T (filled
symbols) already reported in Fig.~\ref{fig:1} compared to
theoretical curves. The solid line superimposed to the upper curve
is a best-fitting BTK function (see eq.\ref{eq:1}). The solid line
superimposed to the lower curve is \emph{not} a best-fitting
curve, but the function obtained from the previous one by taking
$\sigma\ped{\pi}=1$ (see eq.\ref{eq:2}). For details see the
text.}} \label{fig:2}\vspace{-3mm}
\end{figure}
Figure~\ref{fig:1} shows the normalized conductance curves of an
$ab$-plane contact, for increasing intensities of the magnetic
field applied parallel to the $c$ axis. The differential
conductance d$I$/d$V$ was numerically calculated from the measured
$I-V$ characteristic of the junction. The normalization was made
by dividing the experimental d$I$/d$V$ vs. $V$ curves by the
linear or quartic function that best fits them for
$|V|\!\!>\!\!30$~meV. The zero-field curve shows two peaks at
$V\simeq \pm 2.7$~mV and $V\!\simeq \pm 7.2$~mV, clearly related
to the two gaps $\Delta\ped{\pi}$ and $\Delta\ped{\sigma}$,
respectively. On increasing the magnetic field intensity, the
peaks related to the smaller gap, $\Delta\ped{\pi}$, are quickly
suppressed, and they finally disappear at $B=\mu\ped{0}H \simeq
1$~T. On the contrary, the features connected to the large gap,
$\Delta\ped{\sigma}$, look practically unchanged. The whole shape
of the conductance curves changes very little if the field
slightly exceeds the value $B=1$~T, suggesting that in this
magnetic-field region the $\sigma$ band is quite robust. On
further increasing the field intensity, the two peaks
corresponding to $\Delta\ped{\sigma}$ decrease in amplitude and
gradually shift to lower energies, i.e. the large gap gradually
closes. At $B=4$~T, however, the normalized conductance is still
far from being flat. This means that, at the liquid helium
temperature, the upper critical field of the $\sigma$ band for
$\mathbf{B}\parallel c$ is greater than 4~Tesla, in contrast to
the value $H\ped{c2}^{\parallel c}=30$~kOe given by recent torque
magnetometry \cite{Angst} and thermal conductivity measurements
\cite{Sologubenko}. The possible reason for this discrepancy will
be discussed later on.

Let us discuss for a while the quick suppression of the small-gap
features and their disappearance at $B$=1~T. This result was
already obtained by point-contact measurements in polycrystalline
samples \cite{Szabo}, and was interpreted as due to the selective
suppression of the superconductivity in the $\pi$ bands. In our
case, the use of single crystals allows giving more convincing
arguments to support this interpretation, i.e. to show that the
magnetic field of 1~Tesla really destroys the superconductivity
in the $\pi$ bands, without affecting the $\sigma$ bands.

\begin{figure*}[ht]
\begin{center}
\includegraphics[keepaspectratio,width=0.9\textwidth]{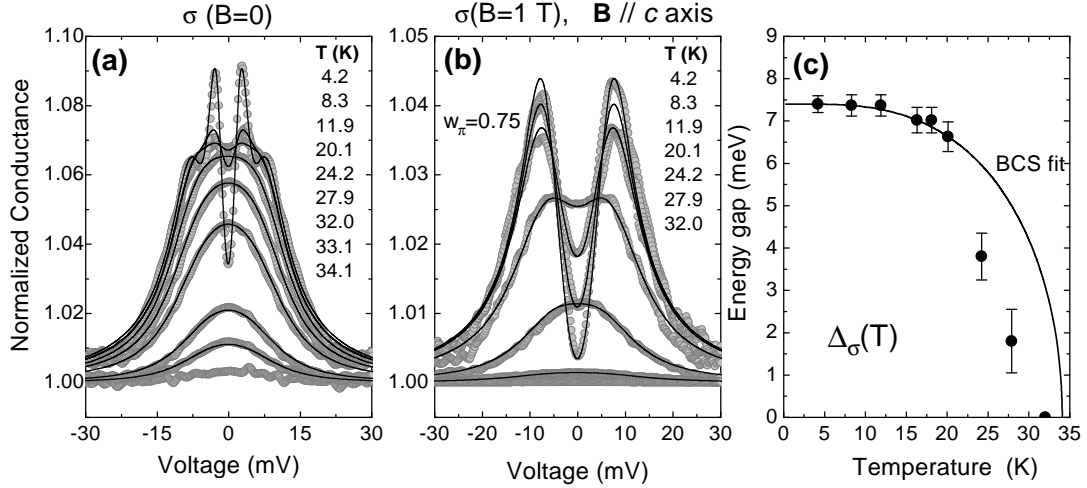}
\end{center}
\vspace{-4.5cm} \caption{\small{(a) Temperature dependence of the
experimental normalized conductance of a $ab$-plane Ag contact
($R\ped{N}=20\, \Omega$) . Solid lines are the best-fitting
functions given by eq.\ref{eq:1}. (b) Same as in (a) but with a
magnetic field of 1 Tesla applied parallel to the $c$ axis. Solid
lines represent the best-fitting curves given by eq.\ref{eq:2}.
(c) Thermal evolution of the large gap $\Delta\ped{\sigma}$ as
obtained from the fit of the curves in (b).}}
\vspace{-3mm}\label{fig:3}
\end{figure*}

As a matter of fact, in Figure \ref{fig:2} the zero-field curve
(open symbols) and the curve in a field of 1 T (filled symbols)
are compared. The solid line superimposed to the former is its
best-fitting curve obtained with the BTK model \cite{BTK}
generalised to the case of two bands, in which the
\emph{normalized} conductance $\sigma$ is expressed by a weighed
sum of the partial BTK conductances of the $\pi$ and $\sigma$
bands:
\begin{equation}
\sigma = w\ped{\pi}\sigma\ped{\pi}+(1
-w\ped{\pi})\sigma\ped{\sigma}. \label{eq:1}
\end{equation}
In practice, the total conductance across the junction is thought
of as the parallel of two (independent) channels.  Here
$w\ped{\pi}$ is the weight of the channel connected to the $\pi$
band, i.e. the partial contribution of the $\pi$ band to the
total conductance. $w\ped{\pi}$ is a function of the plasma
frequencies $\omega\ped{p}^{\sigma}$ and $\omega\ped{p}^{\pi}$,
that are much different in the $ab$-plane and along the $c$ axis.
As a result, $w\ped{\pi}$ depends on the angle $\varphi$ between
the direction of current injection and the boron planes
\cite{Brinkman}. For current injection purely along the $ab$
plane, the value $w\ped{\pi}=0.66$ is theoretically predicted
\cite{Brinkman}.

The fitting function expressed by eq.~\ref{eq:1} contains 7
adjustable parameters ($\Delta\ped{\sigma}$ and $\Delta\ped{\pi}$,
the broadening parameters $\Gamma\ped{\sigma}$ and
$\Gamma\ped{\pi}$, the barrier transparency coefficients
$Z\ped{\sigma}$ and $Z\ped{\pi}$, plus the weight factor
$w\ped{\pi}$) so there is some uncertainty in the choice of the
best-fitting values, especially as far as $Z\ped{\sigma,\pi}$ and
$\Gamma\ped{\sigma, \pi}$ are concerned. The curve superimposed to
the zero-field conductance in Fig.~\ref{fig:2} was obtained by
using: $\Delta\ped{\pi}=2.8$ meV, $\Delta\ped{\sigma}=7.2$ meV,
$Z\ped{\pi}=0.48$, $Z\ped{\sigma}=0.94$, $\Gamma\ped{\pi}=1.49$
meV, $\Gamma\ped{\sigma}=3.3$ meV, and finally $w\ped{\pi}=0.75$.
The gap amplitudes agree very well with those predicted by the
two-band model \cite{Brinkman}. The disagreement between the
predicted value of $w\ped{\pi}$ (0.66) and that given by the fit
(0.75) is simply due to the fact that, as previously pointed out,
the current is injected within a solid angle rather than along a
precise direction. As a matter of fact, it can be shown that the
value $w\ped{\pi}=0.75$ is compatible with an injection cone about
26$^{\circ}$ wide \cite{PRLnostro}.

If the superconductivity in the $\pi$ band is destroyed without
affecting the $\sigma$ band, the total conductance across the
junction is expressed by:
\begin{equation}
\sigma = w\ped{\pi}+(1 -w\ped{\pi})\sigma\ped{\sigma}. \label{eq:2}
\end{equation}
which is obtained by taking $\sigma\ped{\pi}$=1 in eq.\ref{eq:1},
and thus only contains the free parameters $\Delta\ped{\sigma}$,
$\Gamma\ped{\sigma}$ and $Z\ped{\sigma}$, plus the weight factor
$w\ped{\pi}$. This is indeed the functional form of the curve
shown in Fig.\ref{fig:2}, superimposed to the conductance curve at
$B=1$~Tesla. All the parameters are unchanged with respect to the
zero-field curve, apart from the barrier parameter that was set to
$Z\ped{\sigma}=0.56$ to adjust the height of the theoretical
curve. It is clear that this function reproduces both the position
of the conductance peaks and the shape of the conductance well
around zero bias. This demonstrates that a field of 1 Tesla
parallel to the $c$ axis really destroys the superconductivity in
the $\pi$ band, without affecting the amplitude of the
$\sigma$-band gap.

One can now wonder whether this is true also when the temperature
is increased, or rather a temperature $T^{*}<T\ped{c}$ exists at
which also the $\sigma$ bands start being affected by the field.
Fig.~\ref{fig:3} reports the temperature dependence of the
experimental curves of the same $ab$-plane contact discussed so
far, in zero field (a) and in the presence of a field of 1~Tesla
parallel to the $c$ axis (b). Even at a first sight, it is clear
that the curves in zero field become almost flat at $T=34.1$~K, --
which is therefore close to the critical temperature of the
junction -- while the curves in the presence of the field -- that
only contain the $\sigma$-band conductance -- flatten at $T=32$~K.
This suggests by itself that the magnetic field causes the
transition to the normal state at a temperature lower than
$T\ped{c}$. This conclusion can be further supported by extracting
the temperature dependence of $\Delta\ped{\sigma}$ from the
conductance curves in Fig. \ref{fig:3}b, which requires fitting
the experimental curves. The solid lines superimposed to the
experimental data in (a) and (b) represent the relevant
best-fitting curves obtained by using eq.\ref{eq:1} and
eq.\ref{eq:2}, respectively. In both cases, the weight
$w\ped{\pi}$ was taken as temperature-independent. The barrier
parameters $Z\ped{\sigma}$ and $Z\ped{\pi}$ were kept (almost)
constant at the increase of the temperature. Instead, the
broadening parameters $\Gamma\ped{\pi}$ and $\Gamma\ped{\sigma}$
given by the fit increase with $T$. In the case of
Fig.~\ref{fig:3}a, $\Gamma\ped{\pi}$ varies between 1.49 and 2.29
meV, and $\Gamma\ped{\sigma}$ increases from 3.3 up to 3.6 meV;
$Z\ped{\sigma}$ and $Z\ped{\pi}$ vary from 0.94 to 0.8 and from
0.48 to 0.33, respectively. In the case of Fig.~\ref{fig:3}b,
instead, $\Gamma\ped{\sigma}$ is equal to 3.7 meV at low $T$ and
increases rapidly on heating, while $Z\ped{\sigma}$ remain in the
range between 0.6 and 0.45.

The thermal evolution of $\Delta\ped{\sigma}$, obtained by fitting
the curves in Fig.~\ref{fig:3}b with the three-parameter function
given by eq.~\ref{eq:2}, is reported in Fig.~\ref{fig:3}c
(circles) together with a BCS-like curve (solid line). It is clear
that the points sudden deviate from the BCS behaviour at
$T^{*}\sim 20$~K. Based on previous determinations of the
temperature dependence of $\Delta\ped{\sigma}$ in zero field
\cite{PRLnostro}, we can rather safely assume that such a large
deviation is due to the progressive closing of the large gap due
to the magnetic field. At $T=32$~K, the $\sigma$-band gap
disappears: this means that, at this temperature,
$B\ped{c2}^{\parallel c} \sim 1$~T. Again, this value is larger
than that recently measured in the same crystals by other groups
\cite{Angst,Sologubenko}.

\subsection{Magnetic field parallel to the $ab$ planes}
\vspace{3pt}
\begin{figure}[t]
\begin{center}
\includegraphics[keepaspectratio,width=0.9\columnwidth]{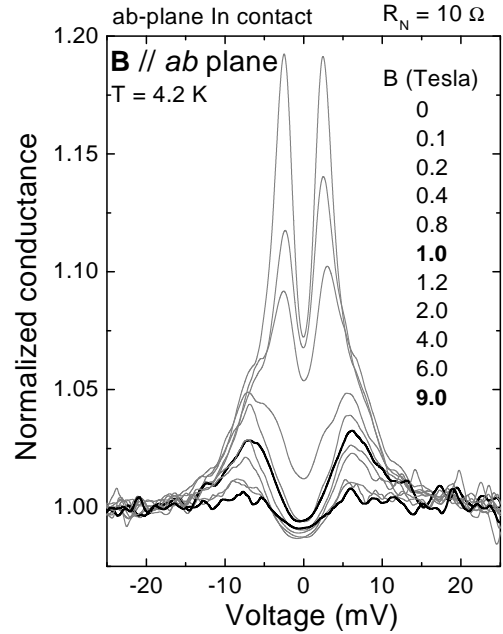}
\end{center}
\vspace{-2.3cm} \caption{\small{Magnetic-field dependence of the
low-temperature normalized conductance curves of a $ab$-plane In
contact. The magnetic field is applied parallel to the $ab$ plane.
Thick lines represent the curves at 1 T and 9 T. }}\label{fig:4}
\end{figure}

\begin{figure*}[t]
\begin{center}
\includegraphics[keepaspectratio,width=0.9\textwidth]{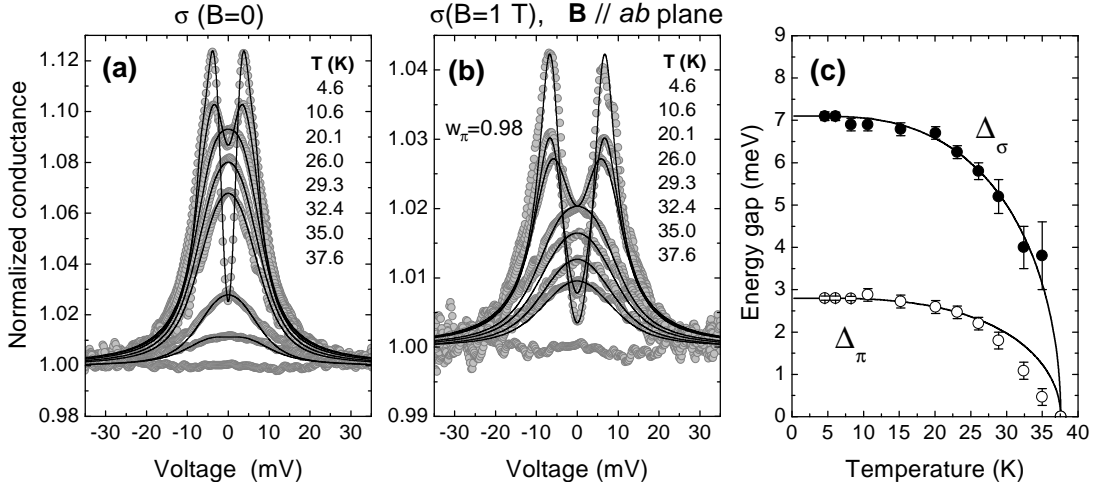}
\end{center}
\vspace{-4.3cm} \caption{\small{(a) Temperature dependence of the
experimental normalized conductance of a $c$-axis In contact
($R\ped{N}=50\,\Omega$). Solid lines are the best-fitting curves
of the form given in eq.\ref{eq:1}. (b) Same as in (a) but in a
magnetic field of 1 Tesla applied parallel to the $ab$ plane.
Lines: best-fitting curves expressed by eq.\ref{eq:2}. (c)
Temperature dependence of the large gap $\Delta\ped{\sigma}$ from
the fit of the curves in (b), and of the small gap
$\Delta\ped{\pi}$ from the fit of the difference between the
curves in (a) and (b) (see ref.\cite{PRLnostro} for details).}}
\vspace{-3mm}\label{fig:5}
\end{figure*}

Fig.~\ref{fig:4} shows the conductance curves of a $ab$-plane In
contact as a function of the intensity of the magnetic field,
applied parallel to the $ab$ planes. Exactly as in
Fig.\ref{fig:1}, the small-gap features are very easily disrupted
by the field. At $B$=1~T, $\Delta\ped{\pi}$ completely vanishes
while $\Delta\ped{\sigma}$ remains unchanged. Based on arguments
similar to those used in the $\mathbf{B}\parallel c$ case, we can
conclude that, at the liquid helium temperature, the upper
critical field of the $\pi$ bands is isotropic, as shown in
ref.~\cite{Bouquet}, and takes the value $B\ped{c2}^{\pi}= \mu
H\ped{c2}^{\pi}\simeq 1$~T. Contrary to what happens in a magnetic
field applied parallel to the $c$ axis, the conductance peaks
associated to the $\sigma$-band gap remain well distinguishable up
to 9 T. A small decrease in the gap amplitude is observed above 4
T, together with an increase in the zero-bias conductance. It
looks thus clear that, at 4.2 K, the critical field of the
$\sigma$ band in the $\mathbf{B}\parallel ab$ case is much greater
than 9 T, in agreement with other experimental findings.

As already done in the $\mathbf{B}\parallel c$ case, let us check
whether the robustness of the $\sigma$ bands upon application of a
field of 1 Tesla persists at the increase of the temperature.
Figure \ref{fig:5} reports the temperature evolution of the
normalized conductance curves (symbols) in zero field (a), and in
the presence of a field of 1 Tesla (b) parallel to the $ab$ plane.
It is clearly seen that, in this case, all the curves become flat
at the same temperature $T=37.6$~K -- which is the critical
temperature of the junction -- indicating that the
superconductivity in the $\sigma$ band survives up to $T\ped{c}$
even in the presence of the field. Solid lines in
Figs.~\ref{fig:5}a and \ref{fig:5}b represent the best-fitting
curves obtained by using eqs.\ref{eq:1} and \ref{eq:2},
respectively. The value of the weight $w\ped{\pi}=0.98$ was
determined by fitting the zero-field, low-temperature curve and
then kept constant \footnote{For current injection purely along
the $ab$ planes the predicted value of the weight is
$w\ped{\pi}=0.99$. Our value is compatible with a current
injection cone of about $60^{\circ}$.}. Let us disregard the
values of the fitting parameters of the zero-field curves, as they
are not essential for our reasoning. As far as the curves in
Fig.~\ref{fig:5}b are concerned, the low-temperature values of the
fitting parameters are: $\Delta\ped{\pi}=7.1$ meV,
$\Gamma\ped{\sigma}=1.75$ meV and $Z\ped{\sigma}=0.58$. At the
increase of the temperature, $\Gamma\ped{\sigma}$ regularly
increases up to 4.2 meV close to $T\ped{c}$, while $Z$ slightly
decreases down to 0.35. The temperature dependence of
$\Delta\ped{\sigma}$ obtained by the fit is reported in
Fig.~\ref{fig:5}c (solid symbols) and compared to a BCS-like curve
(solid line). In agreement with previous experimental findings and
theoretical predictions \cite{Liu,Brinkman}, the large gap is
found to follow rather well the BCS curve. Incidentally, this
confirms \emph{a posteriori} that the $\sigma$-band gap is not
affected by the field of 1 Tesla even at temperatures rather close
to $T\ped{c}$. Also notice the much smaller error on the gap
values with respect to measurements in polycrystalline samples,
due to the reduction in the number of adjustable parameters (from
6 to 3) obtained by removing the $\pi$ band gap by means of the
field \cite{PRLnostro}.

An highly accurate determination of the temperature evolution of
$\Delta\ped{\pi}$ is possible as well, by subtracting from the
total conductance the partial contribution of the $\sigma$ band.
In practice, each experimental curve in Fig.~\ref{fig:5}b is
subtracted from the curve in zero field measured at the same
temperature, as discussed elsewhere \cite{PRLnostro}, and the
resulting curve is fitted by a function of the form $\sigma =
w\ped{\pi}(\sigma\ped{\pi}-1)$, with $w\ped{\pi}=0.98$. The
temperature dependence of $\Delta\ped{\pi}$ obtained from the fit
is shown in Fig.~\ref{fig:5}c (open symbols) and compared to a
BCS-like behaviour (solid line). At $T\gtrsim 20$~K,
$\Delta\ped{\pi}$ deviates from the BCS curve of an amount which
is well greater than our experimental uncertainty. This deviation
is indeed predicted by the two-band model \cite{Liu,Brinkman} but
its unquestionable determination was impossible so far, because of
the very large error affecting the gap values near $T\ped{c}$.
\begin{figure}[t]
%\vspace{-8mm}
\begin{center}
\includegraphics[keepaspectratio,width=0.9\columnwidth]{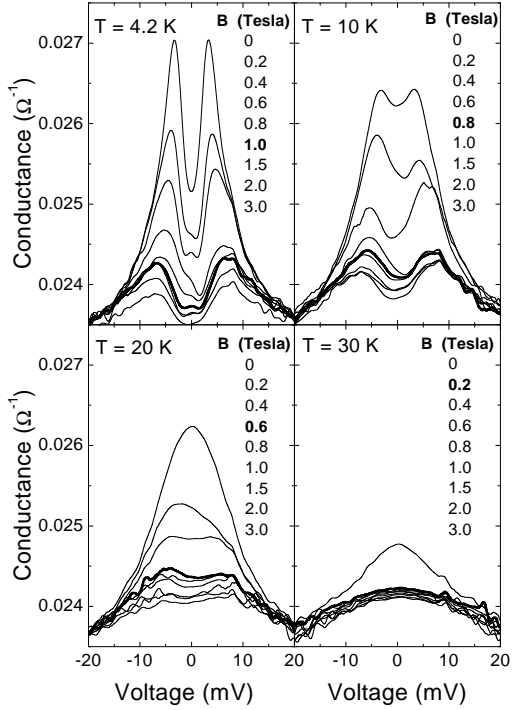}
\end{center}
\vspace{-10mm} \caption{\small{Unnormalized conductance curves of
a $c$-axis In contact ($R\ped{N}\sim 42\,\Omega$) in a magnetic
field of increasing intensity, applied parallel to the $ab$ plane.
Each panel refers to a different temperature, indicated in the
graph. }}\label{fig:6}
\end{figure}

\begin{figure}[ht]
\begin{center}
\includegraphics[keepaspectratio,width=0.85\columnwidth]{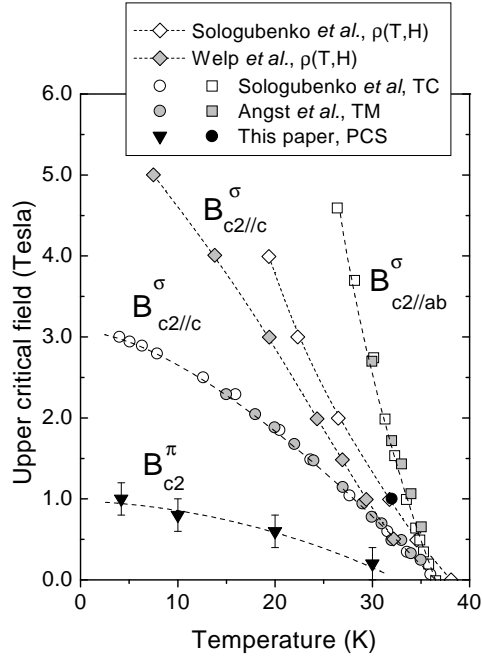}
\end{center}
\vspace{-13mm} \caption{\small{Phase diagram of MgB$\ped{2}$ as it
results from different kinds of measurements (TM = torque
magnetometry, TC = thermal conductivity, PCS = point-contact
spectroscopy, $\rho(T,H)$ = resistivity) carried out in single
crystals. Lines are only guides to the eye. }}\label{fig:7}
\vspace{-0.3cm}
\end{figure}

\subsection{Temperature dependence of $B\ped{c2}^{\pi}$}
\vspace{3mm}

Finally, let us determine the temperature evolution of the
critical field of the $\pi$ band, whose value at 4.2 K has been
evaluated by analyzing the curves in Figs.~\ref{fig:1} and
\ref{fig:4}. Fig.~\ref{fig:6} reports the unnormalized (i.e.,
as-measured) conductance curves of a $c$-axis contact, in the
presence of a magnetic field applied parallel to the $ab$ planes,
at four temperatures: 4.2, 10, 20 and 30 K. Based on our previous
finding that $B\ped{c2
\parallel ab}^{\pi}=B\ped{c2\parallel c}^{\pi}=1$ T at 4.2 K, and on the
essentially isotropic character of the $\pi$ bands
\cite{Kortus,Shulga,An,Kong,Bouquet} we will assume that the above
equality holds at any temperature. The use of a $c$-axis contact
emphasizes the $\pi$-band contribution to the conductance, but
makes it more difficult to distinguish the (very small)
$\sigma$-band features that survive after the removal of
$\Delta\ped{\pi}$. However, we expect that, just above
$B\ped{c2}^{\pi}$, the conductance curves should be (relatively)
field-independent, as long as the residual $\sigma$-band gap is
not suppressed. The curve that marks the beginning of this
``saturation'' is represented by a thick curve in the four panels
of Fig.\ref{fig:6} \footnote{Notice that only the curves at some
representative field intensities are reported for clarity.}. At
4.2, 10 and 20 K a residual gap-like feature is clearly observed
in these curves, thus indicating the persistence of
superconductivity in the $\sigma$ bands. At $T=30$ K the
saturation occurs at $B\simeq 0.2$ T and above this field only
minor changes in the conductance are observed. Since the upper
critical field of MgB$\ped{2}$ in the $\mathbf{B}\parallel ab$
case is, even at this temperature, at least one order of magnitude
greater than 0.2~T \cite{Angst,Sologubenko,Welp}, there is no
doubt that this saturation is due to the removal of the $\pi$-band
gap alone.

The values of the magnetic field intensities that give rise to
this saturation, and that we interpret as $B\ped{c2}^{\pi}$, are
reported in Fig.\ref{fig:7} (black triangles) together with the
results of torque magnetometry \cite{Angst}, resistivity
\cite{Sologubenko,Welp} and thermal conductivity measurements
\cite{Sologubenko}. There are two points that are worth
mentioning: the clear overestimation of $B\ped{c2\parallel
c}^{\sigma}$ by transport measurements, and the fact that all the
critical fields determined by bulk measurements (e.g. thermal
conductivity and torque magnetometry) vanish at a temperature
$T<T\ped{c}$, where $T\ped{c}$ is determined by the resistive
transition. These puzzling results have been recently interpreted
as due to the existence of an additional phase with enhanced
critical parameters ($T\ped{c}$ and $H\ped{c2}$), probably related
to surface effects that seem to be strongly suppressed by in-plane
magnetic fields \cite{Welp}. If this is the case, the results of
surface-sensitive measurements might be sensibly different from
those of bulk-sensitive techniques. This is an important point in
discussing our results, since PCS is intrinsically a
surface-sensitive probe. As a matter of fact, the value we
determined for $B\ped{c2\parallel c}^{\sigma}$ (black circle) is
well compatible with the results of resistive measurements by
Sologubenko \emph{et al.}, that were carried out on similar single
crystals. Moreover, the observation of $\sigma$-band
superconductivity at 4.2 K in a field of 4 T parallel to the $c$
axis (see fig.\ref{fig:1}) further supports this picture.

In conclusion, we have presented the results of a systematic
study of MgB$\ped{2}$ by means of point-contact spectroscopy in
the presence of a magnetic field. The use of single crystals has
allowed us to control the direction of both the current injection
and the applied magnetic field. Consequently, we have been able
to study the effect of the magnetic field on each band separately,
and to determine for the first time the temperature dependence of
the upper critical field of the isotropic $\pi$ bands. As far as
the $\sigma$ bands are concerned, the effect of the magnetic
field has been confirmed to be strongly anisotropic. The obtained
value of the upper critical field $B\ped{c2 \parallel c}^{\sigma}$
is higher than that measured on the same crystals by torque
magnetometry and thermal conductivity, but agrees very well with
the results of transport measurements. Finally, by exploiting the
directionality of the point contacts and the different effect of
the magnetic field on the two band systems, we have selectively
destroyed the superconductivity in the $\pi$ bands. This procedure
allows separating the partial contributions of the $\sigma$ and
$\pi$ bands to the total conductance of the point contacts.
Fitting each partial conductance with the standard BTK model, we
have determined with great accuracy the temperature dependence of
the two gaps, confirming the predictions of the two-gap models
appeared in literature.

This work was supported by the INFM Project PRA-UMBRA and by the
INTAS project ``Charge transport in metal-diboride thin films and
heterostructures''.


\begin{thebibliography}{10}
%
\footnotesize{
\bibitem{Suhl} H. Suhl, B.T. Matthias, and L.R. Walker, Phys. Rev.
Lett. \textbf{3}, 552 (1959).
%
\bibitem{Liu} A.Y. Liu, I.I. Mazin, and J. Kortus, Phys. Rev. Lett. \textbf{87},
87005
(2001).
%
\bibitem{Brinkman} A. Brinkman \emph{et al.}, Phys. Rev. B \textbf{65},
180517(R) (2001).
%
\bibitem{Mazin} I.I. Mazin \emph{et al.}, cond-mat/0204013.
%
\bibitem{Kortus} J. Kortus \emph{et al.}, Phys. Rev. Lett. \textbf{86}, 4656
(2001).
%
\bibitem{Shulga} S.V. Shulga \emph{et al.}, cond-mat/0103154.
%
\bibitem{An} J.M. An and W.E. Pickett, Phys. Rev. Lett.
\textbf{86}, 4366 (2001).
%
\bibitem{Kong} Y. Kong, O.V. Dolgov, O. Jepsen, and O.K. Andersen,
Phys. Rev. B \textbf{64}, 020501(R) (2001).
%
\bibitem{Buzea} C. Buzea and T. Yamashita, Supercond. Sci.
Technol. \textbf{14}, R115 (2001).
%
\bibitem{Angst} M. Angst \emph{et al.}, Phys. Rev. Lett.
\textbf{88}, 167004 (2002).
%
\bibitem{Sologubenko} A.V. Sologubenko \emph{et al.}, Phys. Rev. B \textbf{66}, 014504 (2002).
%
\bibitem{ballistic} A.M. Duif, A.G.M. Jansen, and P. Wyder,
J. Phys.: Condens. Matter \textbf{1}, 3157 (1989).
%
\bibitem{Tanaka} S. Kashiwaya and Y, Tanaka, Rep. Prog. Phys.
\textbf{63}, 1641 (2000).
%
\bibitem{Szabo} P. Szab\'{o} \emph{et al.}, Phys. Rev.
Lett. \textbf{87}, 137005 (2001).
%
\bibitem{BTK} G.E. Blonder, M. Tinkham, and T.M. Klapwijk, Phys.
Rev. B \textbf{25}, 4515 (1982).
%
\bibitem{PRLnostro} R.S. Gonnelli \emph{et al.}, cond-mat/0208060.
%
\bibitem{Bouquet} F. Bouquet \emph{et al.}, cond-mat/0207141.
%
\bibitem{Welp} U. Welp \emph{et al.}, cond-mat/0203337.}
\end{thebibliography}
\end{document}